\documentclass[aps,pra,twocolumn]{revtex4}

\usepackage{amsmath}
\usepackage{latexsym}
\usepackage{graphicx}
\usepackage{dcolumn}
\usepackage{color}
\usepackage{bm}
\usepackage{bbold}
\usepackage{epstopdf}
\usepackage{verbatim}

\begin{document}
\title{An Experimental Test of Envariance}
\author{L. Vermeyden$^1$, X. Ma$^1$, J. Lavoie$^2$, M. Bonsma$^1$, U. Sinha$^{1,3}$, R. Laflamme$^{1,4}$, and K.J. Resch$^1$}
\affiliation{$^1$Institute for Quantum Computing and
Department of Physics \& Astronomy, University of Waterloo,
Waterloo, Ontario N2L 3G1, Canada}
\affiliation{$^2$Group of Applied Physics, University of Geneva, CH-1211 Geneva 4, Switzerland}
\affiliation{$^3$Raman Research Institute, C. V. Raman Avenue, Sadashivanagar, Bangalore 560 080, India}
\affiliation{$^4$Perimeter Institute for Theoretical Physics, 31 Caroline Street North, Waterloo, Ontario N2L 2Y5, Canada}

\begin{abstract}
\noindent Envariance, or environment-assisted invariance, is a
recently identified symmetry for maximally entangled states in quantum theory with important
ramifications for quantum measurement, specifically for understanding Born's rule \cite{zurek03.1,zurek03.2,zurek05}. We benchmark the degree to
which nature respects this symmetry by using entangled photon
pairs. Our results show quantum states can be
$(99.66\pm0.04)\%$ envariant as measured using the quantum
fidelity \cite{jozsa94}, and $(99.963\pm0.005)\%$ as measured
using a modified Bhattacharya Coefficient
\cite{bhattacharyya43}, as compared with a perfectly envariant
system which would be 100\% in either measure. The deviations can be
understood by the less-than-maximal entanglement in our photon
pairs.
\end{abstract}

\maketitle

Symmetries play a central role in physics with wide-reaching
implications in fields as diverse as spectroscopy and particle
physics. It is therefore of fundamental importance to identify
and understand new symmetries of nature. One of these more recently
identified symmetrys in quantum mechanics has been named
environment-assisted invariance, or \emph{envariance}
\cite{zurek03.1}. It applies in certain cases where a composite
quantum object consists of a system part, labelled $\mathsf{S}$, and an
environment part, labelled $\mathsf{E}$.  If some action is applied to the
system part only, described by some unitary transformation,
$U_\mathsf{S}$, then the state is said to be envariant under
$U_\mathsf{S}$ if another unitary applied to the environment,
$U_\mathsf{E}$, can restore the initial state. This can be expressed,
\begin{eqnarray}
U_\mathsf{S} |\psi_{\mathsf{SE}} \rangle &=& (u_\mathsf{S} \otimes \mathbb{1}_\mathsf{E})|\psi_{\mathsf{SE}}\rangle=|\eta_{\mathsf{SE}} \rangle \\
U_\mathsf{E} |\eta_{\mathsf{SE}} \rangle &=& (\mathbb{1}_\mathsf{S} \otimes u_\mathsf{E})|\eta_{\mathsf{SE}}\rangle = |\psi_{\mathsf{SE}} \rangle.
\end{eqnarray}
\noindent Envariance is an example of an assisted symmetry
\cite{zurek05} where once the system is transformed under some
unitary $U_\mathsf{S}$, it can be restored to its original
state by another operation on a physically distinct system: the
environment.

Envariance is a uniquely quantum symmetry in the following
sense. A pure quantum state represents complete knowledge of
the quantum system. In an entangled quantum state, however,
complete knowledge of the whole system does not imply complete
knowledge of its parts. It is therefore possible that an
operation on one part of a quantum state can alter the global
state, but its local effects are masked by incomplete knowledge of that part; the
effect on the global state can then be undone by an action on a
different part. In contrast, complete knowledge of a composite
classical system implies complete knowledge of each of its
parts. Thus transforming one part of a classical system cannot
be masked by incomplete knowledge and cannot be undone by a
change on another part.

Envariance plays a prominent role in work related to fundamental
issues of decoherence and quantum measurement
\cite{zurek03.1,zurek03.2,zurek05}. Decoherence converts
amplitudes in coherent superposition states to probabilities in
mixtures and is central to the emergence of the
classical world from quantum mechanics
\cite{zurek91,schlosshauer05}. Mathematically the mixture
appears in the reduced density operator of the system which is
extracted from the global wavefunction by a partial trace
\cite{landau27,nielsen00}. This partial trace limits the
approach for deriving, as opposed to separately postulating,
the connection between the wavefunction and measurement
probabilities known as Born's rule \cite{born26}, since the
partial trace \emph{assumes} Born's rule is valid
\cite{zurek03.1,schlosshauer03}.  Envariance was employed in a
derivation of Born's rule which sought to avoid circularity
inherent to approaches which rely on partial trace
\cite{zurek03.1}.  For comments on this
derivation, see for example
\cite{schlosshauer03,barnum03,mohrhoff04}.

In the present work, we subject envariance to experimental test
in an optical system. We use the polarization of a single
photon to encode the system, $\mathsf{S}$, and the polarization
of a second single photon to encode the environment,
$\mathsf{E}$.
We subject the system photon to a wide range of
polarization rotations with the goal of benchmarking the degree
to which we can restore the initial state by applying a second
transformation on the environment photon.

\begin{figure}
  \centering
  \includegraphics[scale=1,width=1\columnwidth]{{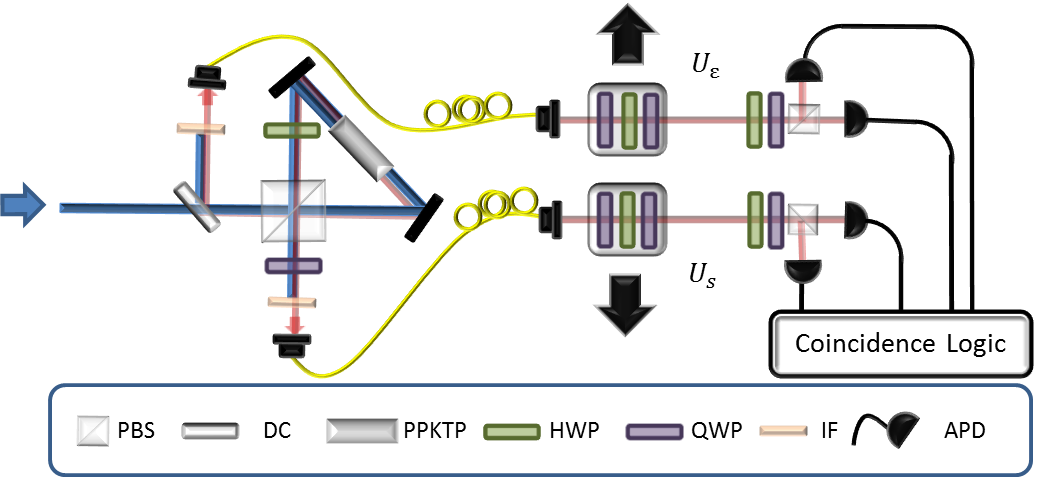}}
  \caption{\label{fig:experimentalsetup} Experimental setup. The entangled
  photon pairs are created using type-II spontaneous parametric down-conversion.
  The pump laser is focused on a periodically-poled KTP crystal and pairs of
  entangled photons with anti-correlated polarizations are emitted. The pump
  is filtered using a band-pass filter, and polarization controls adjust for
  the alterations due to the coupling fibres. The entangled photon pairs are set
  so one photon is considered the system, and the other is considered the
  environment. After the source the unitary transformations are applied. A
  three wave plate combination is required to apply an arbitrary unitary transformation:
  quarter wave-plate (QWP), half wave-plate (HWP), QWP. A set of this combination of
  wave plates is mounted on each translation stage which can slide the wave plates
  in and out of the path of the incoming photons. The photons are then detected using
  polarizing beam splitters (PBS) and two wave plates to take projective
  measurements. The counts are then analyzed using coincidence logic.}\label{fig:setup}
\end{figure}

Our test requires a source of high-quality two-photon
polarization entanglement, an optical set-up to perform unitary
operations on zero, one, or both of the photons, and
polarization analyzers to characterize the final state of the
light. Our experimental setup is shown in Fig.~\ref{fig:setup}.
We produce pairs of polarization-entangled photons using
spontaneous parametric down-conversion (SPDC) in a Sagnac
interferometer \cite{kim06,fedrizzi07,biggerstaff09}. In the
ideal case, this source produces pairs of photons in the  singlet state,
\begin{equation}
|\psi_\mathsf{SE}\rangle=\frac{1}{\sqrt{2}}\left(|H\rangle_\mathsf{S}|V\rangle_\mathsf{E}-|V\rangle_\mathsf{S}|H\rangle_\mathsf{E}\right),
\end{equation}
where $|H\rangle$ ($|V\rangle$) represents horizontal
(vertical) polarization, and $\mathsf{S}$ and $\mathsf{E}$ label
the photons. This state  is envariant under all unitary transformations
and has the convenient symmetry that
$u_\mathsf{S} = u_\mathsf{E}$ for all $u_\mathsf{S}$.
 We pump a $10$~mm periodically-poled KTP crystal (PPKTP),
phase-matched to produce photon pairs at $809.8$~nm and
$809.3$~nm from type-II down-conversion using $6$~mW from a CW
diode pump laser with centre wavelength $404.8$~nm. The output
from the source is coupled into single-mode fibres, where polarization
 controllers correct unwanted polarization
rotations in the fibre. The light is coupled out of the fibres and
directed to two independent polarization analyzers. Each
analyzer consists of a half-wave plate (HWP), quarter-wave
plate (QWP), and a polarizing beam-splitter (PBS). Between the
fibre and the analyzers are two sets of wave plates---a QWP, a
HWP, then another QWP---which can be inserted as a group into the beam
paths to implement controlled polarization transformations.
Photons from both ports of each PBS are detected using
single-photon counting modules (Perkin-Elmer SPCM-AQ4C) and
analyzed using coincidence logic with a $1$~ns coincidence
window, counting for $5$~s. We typically measured total
coincidence rates of $5.4$~kHz across the four
detection possibilities for photons $\mathsf{S}$ and
$\mathsf{E}$.

\begin{table}[t!]
\centering{
\begin{tabular}{c c c c}
\hline
\\ [-1.5ex]
Rotation Axis & $\alpha(\theta)$ & $\beta(\theta)$ & $\gamma(\theta)$  \\ [.5ex]
\hline
$\hat{x}$ & $\pi/2$ & $-\theta/4$ & $\pi/2$ \\
$\hat{y}$ & $\pi/2 + \theta/2$ & $\theta/4$ & $\pi/2$ \\
$\hat{z}$ & $\pi/4$ & $-\pi/4 -\theta/4$ & $\pi/4$ \\
\hline
\end{tabular}}
\caption{Wave plate settings used to implement polarization
rotations. The angles $\alpha$, $\beta$ and $\gamma$ are the
wave plate angles for the first QWP, the HWP and the second QWP
respectively. The angle $\theta$ is the rotation angle of the
polarization about the specified axis on the Bloch sphere.}
\label{table:unitaryangles}
\end{table}

For our experiment, we implemented rotations about the standard
$\hat{x}$, $\hat{y}$, and $\hat{z}$ axes of the Bloch sphere;
in addition we implemented rotations about an axis $\hat{m} =
(\hat{x}+\hat{y}+\hat{z})/\sqrt{3}$.  The wave plate angles
used to implement rotations by an angle $\theta$ about the
$\hat{x}$, $\hat{y}$, and $\hat{z}$ axes are shown in Table
\ref{table:unitaryangles}; the angles to implement rotations
about $\hat{m}$ were determined numerically using
\textsc{Mathematica}.

\begin{figure}[t!]
  \centering
  \includegraphics[scale=1,width=1\columnwidth]{{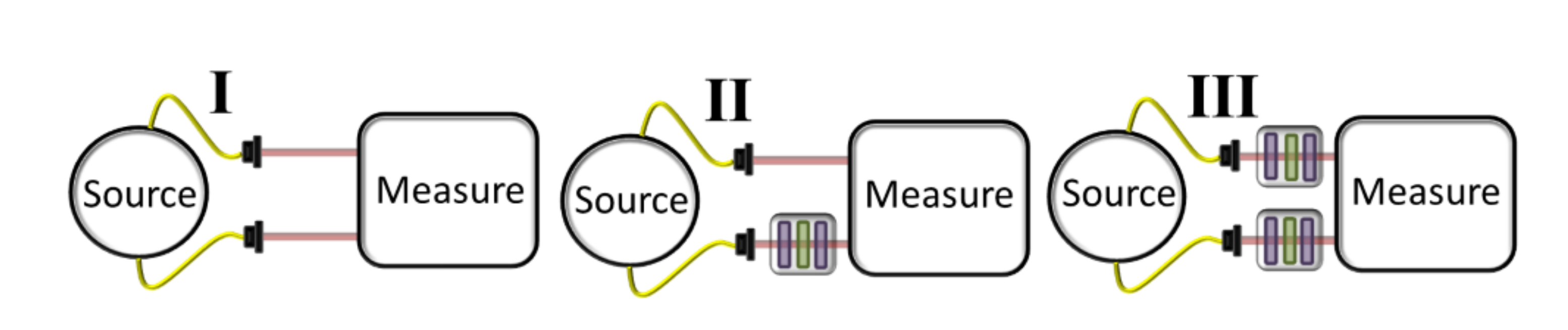}}
  \caption{\label{fig:tanglecalcs} Experimental measurement procedure.  We investigated the impact
  of each unitary transformation by performing quantum state tomography at three different stages:
  directly on the initial state with no unitary transformations \textbf{(I)}, on the state with a transformation applied to the system
  photon \textbf{(II)}, and on a state with the same transformation applied to both the system and environment photon \textbf{(III)}.
}\label{fig:procedure}
\end{figure}

Our experiment proceeds in three stages as depicted in
Fig.~\ref{fig:procedure}: first characterizing the initial
state (\textbf{I}), then characterizing the state after a
transformation is applied to the system photon (\textbf{II}),
and finally characterizing the state after that same
transformation is applied to both system and environment
(\textbf{III}). We record a tomographically-overcomplete set of
measurements at each stage, performing the 36 combinations of
the polarization measurements, $|H\rangle$, $|V\rangle$,
$|D\rangle$=$(|H\rangle+|V\rangle)/\sqrt{2}$,
$|A\rangle$=$(|H\rangle-|V\rangle)/\sqrt{2}$,
$|R\rangle$=$(|H\rangle+i|V\rangle)/\sqrt{2}$, and
$|L\rangle$=$(|H\rangle-i|V\rangle)/\sqrt{2}$ on each photon
and counting for $5$~s for each setting. The states were then
reconstructed using the maximum likelihood method from
Ref.~\cite{jezek03}. This procedure was repeated for a diverse
range of transformations.  We configured our setup to implement
unitary rotations in multiples of 30$^\circ$ from 0$^\circ$ to
360$^\circ$ about each of the $\hat{x}$, $\hat{y}$, $\hat{z}$,
and $\hat{m}$ axes.  The data acquisition time for this
procedure over the set of 13 rotation angles about each axis
was approximately six hours. The source was realigned before each run
to achieve maximum fidelity with the singlet state from $0.985$ to
$0.990$. 
\begin{figure}[t!]
  \centering
  \includegraphics[scale=1,width=1\columnwidth]{{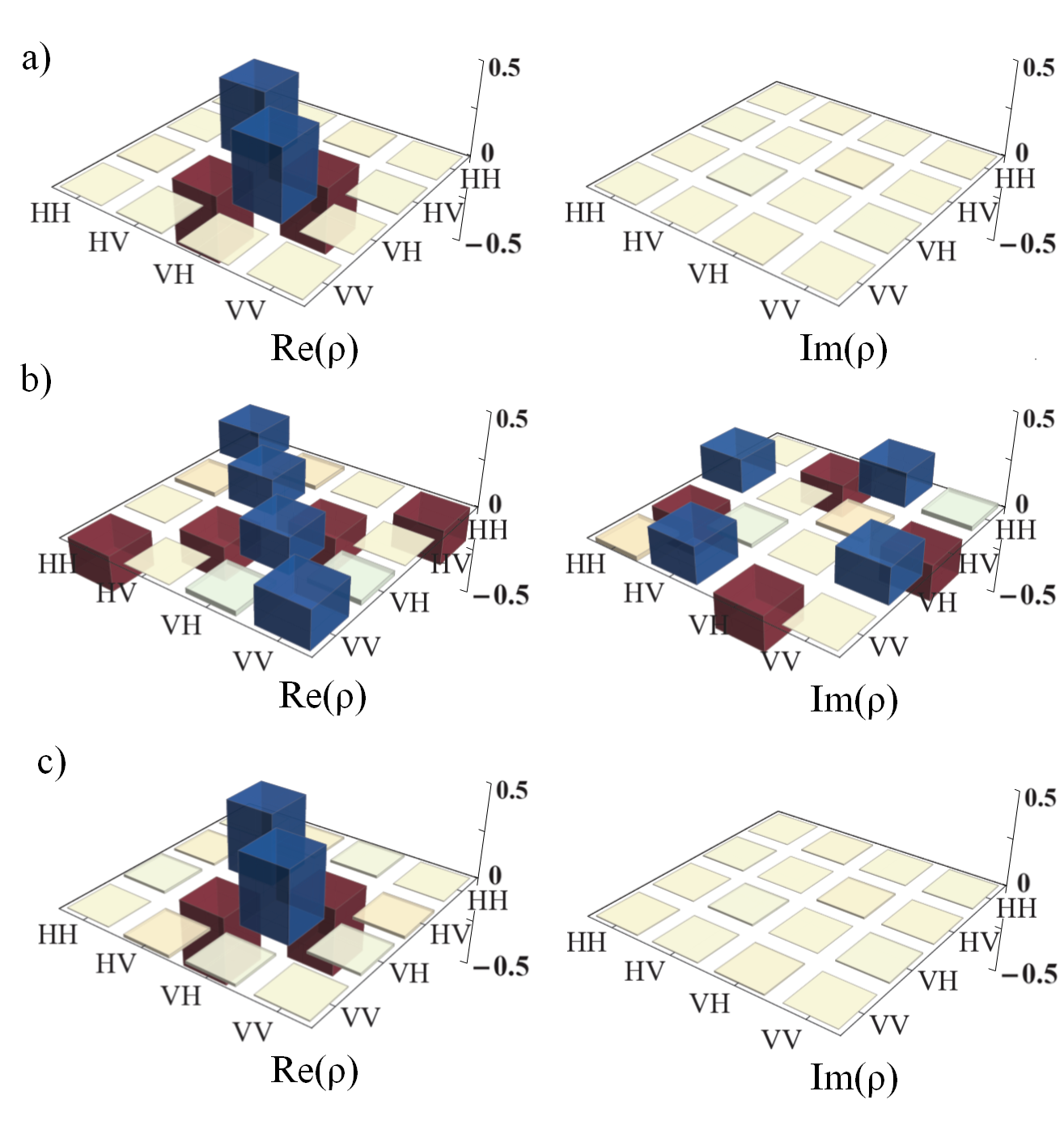}}
  \caption{\label{fig:tomo} a) Is the real and imaginary parts of the reconstructed density matrix of the initial state from the source (stage \textbf{I} of the procedure). It has $0.987$ fidelity \cite{jozsa94} with the ideal. b) The system photon is transformed using wave plates set to implement the rotation of $90^\circ$ about the $\hat{x}$ axis, stage \textbf{II}. The resulting density matrix shown has $0.488$ fidelity with the ideal initial state,$0.501$ with the initial reconstructed state and $0.995$ with the expected state, calculated by transforming the density matrix from a). c) The reconstructed density matrix after the same unitary from b) is applied to both photons, stage \textbf{III}. This state has a $0.987$ fidelity with the ideal, $0.995$ with the reconstructed state from a), and $0.997$ with the expected state calculated by transforming the state from part a).
}\label{fig:tomography}
\end{figure}

Figure ~\ref{fig:tomography}a)--c) show the real and imaginary
parts of the reconstructed density matrix of the quantum state
at the three stages in the experiment, \textbf{I}, \textbf{II},
and \textbf{III} respectively. The fidelity \cite{jozsa94} of
the state with the ideal $\psi^-\rangle$ state during these samples of two of the stages are
$0.987$ for both \textbf{I} a), and \textbf{III} c),
respectively, and is defined as \cite{jozsa94}:
\begin{equation}
F(\rho,\sigma) = {\{\mathrm{Tr}[(\sqrt{\rho}\sigma \sqrt{\rho})^{1/2}]\}}^{2}
\end{equation}
We can use this definition to calculate the fidelity between the state at
stages \textbf{I} and \textbf{III}. Comparing between the states shown in
Fig.~\ref{fig:tomography} panels a) and c) the resulting fidelity is $0.995$.

\begin{figure*}
  \centering
  \includegraphics[scale=1,width=2\columnwidth]{{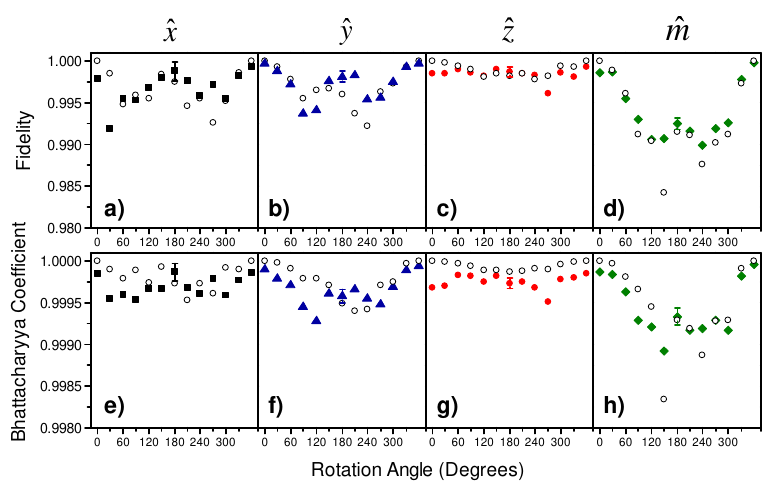}}
  \caption{\label{fig:quantumfidelity}  Analysis of the experimental results. Panels
  a)--d) show the fidelity analysis results for unitary rotations
  about $\hat{x}$, $\hat{y}$, $\hat{z}$, and $\hat{m}$ axes as functions of rotation angle. The coloured data
  points are the comparison between stage \textbf{I} and stage \textbf{III} (comparing the source state and
  the state after the unitary has been applied to both qubits). The open circles show
  a theoretical comparison. Panels e)--h) show the quantum Bhattacharyya results comparing stage \textbf{I}
  and stage \textbf{III} in the coloured data points for each of the four axis, with the open circles being the theoretical
  comparison. For plots which include a comparison of stage \textbf{I} and \textbf{II} (applying the unitary to one qubit only)
  and theoretical comparisons, see the appendix. The error bar for each graph is the
  standard deviation of comparisons of source state measurements during the experiment.}\label{fig:QF}
\end{figure*}

The summary of the results from our experiment is shown in
Fig.~\ref{fig:quantumfidelity}. The coloured data points in
Fig.~\ref{fig:quantumfidelity}a)--d) show the fidelity of the
experimentally reconstructed state at stage \textbf{III} with
the reconstructed state from the initial stage \textbf{I},
i.e.,
$F(\rho_{\rm{expt}}^\mathbf{I},\rho_{\rm{expt}}^\mathbf{III})$,
as a function of the rotation angle for rotations about the
$\hat{x}$, $\hat{y}$, $\hat{z}$, and $\hat{m}$, respectively.
The open circles show the theoretical expectation for the
fidelity between the measured state at stage \textbf{I} with
the expected state in stage \textbf{III}, calculated by acting
the unitaries on the measured state from stage \textbf{I},
i.e.,
$F(\rho_{\rm{expt}}^\mathbf{I},\rho_{\rm{th}}^\mathbf{III})$.
The fidelities are very high, close to the limit of 1, in all
cases and we see reasonable agreement with expectation.

We considered the effects of Poissonian noise and waveplate
calibration on our results and found that these effects were
too small to explain the deviation between
$F(\rho_{\rm{expt}}^\mathbf{I},\rho_{\rm{expt}}^\mathbf{III})$
and
$F(\rho_{\rm{expt}}^\mathbf{I},\rho_{\rm{th}}^\mathbf{III})$.
To account for this, we characterized the fluctuations in the
state produced by the source itself by comparing the state
produced in subsequent stage \textbf{I} sates in the data
collection; recall that stage \textbf{I} for each choice of
unitary is always the same (no additional waveplates inserted)
and thus provides a good measure of the source stability.
Specifically, we calculated the standard deviation in the
fidelity of the state produce at a stage I in the $i^{\rm{th}}$
round of the experiment to that produced in the \emph{next},
$(i+1)^{\rm{th}}$, stage I,
$F(\rho_{\rm{expt}}^{\mathbf{I},i},\rho_{\rm{expt}}^{\mathbf{I},i+1})$.
The standard deviation in these fidelities calculated from the
data taken within each set of rotation axes are shown as
representative error bars on the plots in
Figs.~\ref{fig:quantumfidelity}a)--d). The standard deviation
of this quantity over all the experiments was $0.0008$. We
characterize the difference between the measured and expected
fidelities by calculating the standard deviation in the
quantity,
$F(\rho_{\rm{expt}}^\mathbf{I},\rho_{\rm{expt}}^\mathbf{III})
-F(\rho_{\rm{expt}}^\mathbf{I},\rho_{\rm{th}}^\mathbf{III})$,
for each experiment.  (This is the difference between the
coloured and open data points in
Figs.~\ref{fig:quantumfidelity}a)--d).) over all experiments to
be $0.002$.  This value is comparable to the error in the
fidelity due to source fluctuations. Refer to the appendix to
see the comparison between stage \textbf{I} and stage
\textbf{II}, which would not fit on the scale of
Fig.~\ref{fig:quantumfidelity}.

From our data, we extract the average fidelity
$F(\rho_{\rm{expt}}^\mathbf{I},\rho_{\rm{expt}}^\mathbf{III})$
for the set of measurements made for each unitary axis and show
the results in Table II.  As measured by the average fidelity,
our experiment benchmarks envariance to $0.9966\pm0.0004$,($(99.66\pm0.04)\%$ of the ideal)
averaged over all rotations.

\begin{table}[t!]
\centering{
\begin{tabular}{c c c}
\hline
\\ [-1.5ex]
Rotation Axis & Average Fidelity & Average $BC$ \\ [.5ex]
\hline
$\hat{x}$ & $0.997\pm0.001$ & $0.9997\pm0.0001 $ \\
$\hat{y}$ & $0.9973\pm0.0007$ & $0.99966\pm0.00008$ \\
$\hat{z}$ & $0.9984\pm0.0006$ & $0.99975\pm0.00007$ \\
$\hat{m}$ & $0.9941\pm0.0007$ & $0.9994\pm0.0001$ \\
Overall average: & $0.9966\pm0.0004$ & $0.99963\pm0.00005$ \\
\hline
\end{tabular}}
\caption{Summary of the results for comparing stages \textbf{I} and \textbf{III}
using fidelity and Bhattacharyya Coefficient ($BC$) analysis and averaging over each unitary rotation.
 The overall average is representative of the overall envariance of our state.}
\label{table:fidelitysummary}
\end{table}

Fidelity has conceptual problems as a measure for testing quantum
mechanics, since the density matrix we used to compute the fidelity is reconstructed using state tomography, which is under the assumption of Born's rule. The Bhattacharyya
Coefficient ($BC$) is a measure of the overlap between two discrete distributions $P$ and $Q$, where $p_i$ and $q_i$ are the probabilities of the  $i^{th}$ element for $P$ and $Q$ respectively. The $BC$ is defined \cite{bhattacharyya43},
\begin{equation}
BC = \sum_{i} \sqrt{p_i q_i}.
\end{equation}
If we normalize the measured tomographic data by dividing by
the sum of the counts, we can treat this as a probability
distribution.  The $BC$ then can be
calculated using the distribution of measurements at each stage
in the experiment, directly analogous to the approach used with
fidelity.  It should be noted that the $BC$ has some limitations
when applied in this case. If two quantum states produce
identical measurement outcomes, its value is 1. Unlike fidelity
though, it is not the case that the $BC$ goes to 0 for orthogonal quantum
states. For example, the $BC$ for
two orthogonal Bell states measured with an overcomplete set of
polarization measurements is $7/9$.  Furthermore, the value of
the $BC$ is dependent on the
particular choice of measurements taken. While we are employing
a commonly-used measurement set for characterizing two qubits,
other choices would produce different $BC$s.  Nevertheless, this
metric can be employed to quantify the envariance in our
experiment without quantum assumptions, making it appropriate
for testing quantum mechanics.

The Bhattacharyya Coefficients from our measured data are shown
in Fig.~\ref{fig:quantumfidelity}e)--h).  We normalize the
measured counts from stages \textbf{I} and \textbf{III} to give
us probability distributions $p^\mathbf{I}_{\rm{expt}}$ and
$p^\mathbf{III}_{\rm{expt}}$.  The coloured data points in
Figs.~\ref{fig:quantumfidelity}e)--h) show the $BC$ between
these distributions,
$BC(p_{\rm{expt}}^\mathbf{I},p_{\rm{expt}}^\mathbf{III})$. The
open circles are a theoretical expectation of the $BC$ given
the tomographic measurements from stage \textbf{I}; for these
theoretical values we used state tomography, and thus assumed
quantum mechanics, to obtain the expected distribution
$p_{\rm{th}}^\mathbf{III}$ and calculate the expected $BC$,
$BC(p_{\rm{expt}}^\mathbf{I},p_{\rm{th}}^\mathbf{III})$.

Using an analogous procedure to that employed with the
fidelity, we estimate the uncertainty in the $BC$ by comparing
subsequent measured distributions in stage \textbf{I}
throughout the experiment, i.e.,
$BC(p_{\rm{expt}}^{\mathbf{I},i},p_{\rm{expt}}^{\mathbf{I},i+1})$.
A representative error bar calculated from the data for a set
of unitaries around the same axis are shown in
Fig.~\ref{fig:quantumfidelity}e)--h). The standard deviation in
this quantity over all the data is $0.00005$.  As before we
characterize the difference between the measured and expected
BCs as the standard deviation of the quantity
$BC(p_{\rm{expt}}^\mathbf{I},p_{\rm{expt}}^\mathbf{III})
-BC(p_{\rm{expt}}^\mathbf{I},p_{\rm{th}}^\mathbf{III})$ which
is $0.00009$ over all experiments.  As before, this value is
comparable to the error due to source fluctuations.  Data
showing the $BC$ between stage \textbf{I} and \textbf{II} are
shown in the appendix along with analogous theoretical
comparison. A summary of the $BC$ analysis results are in Table
~\ref{table:fidelitysummary}. The average measured $BC$ is $0.99963\pm0.00005$
($(99.963\pm0.005)\%$ of the ideal) across all tested unitaries.


Our deviation from perfect envariance can be understood from our imperfect state fidelity. 
However, we also consider the magnitude of the violation of Born's rule if one instead assumes all 
of the deviation stems from such a violation.  One recently proposed extension of Born's rule
\cite{wonmin} determines probabilities by raising the wavefunction to the power of $n$ rather than Born's
rule which raises the wavefunction to the power of 2.  In this theory, the correlation between 
measurement outcomes as a function of measurement setting on a singlet state depends on 
the power of $n$, thus we can test this theory using our experimental data.  Fitting our
experimental data to this model, we find $n=2.01 \pm 0.02$ in good agreement with Born's rule.
More details are included in the supplementary materials.

We have experimentally tested the property of envariance on an entangled two-qubit quantum state. Over
a wide range of unitary transformations, we experimentally showed envariance at $(99.66 \pm 0.04)\%$ when
 measured using the fidelity and $(99.963 \pm 0.005)\%$ using the
Bhattacharyya Coefficient. Deviations from perfect envariance are in 
good agreement with theory and can be explained by our initial state fidelity and 
fluctuations in the properties of our state. Fitting our results to a recently published model which does not explictly assume Born's rule yields nevertheless good agreement with it. Our results serve as a benchmark for the property
of envariance, as improving the envariance of the state
signicantly would require substantive improvements in
source delity and stability. It would be interesting to extend tests of envariance to higher dimensional quantum
state and to other physical implementations.

\emph{Acknowledgements}- We thank D. Hamel, and K. Fisher
for valuable discussions.  We are grateful for financial
support from NSERC, QuantumWorks, MRI ERA, Ontario Centres of
Excellence, Industry Canada, Canada Research Chairs, CFI and CIFAR.

\pagebreak

\section{Supplementary Information}

\subsection{Additional Experimental Data}

\begin{figure*}
  \centering
  \includegraphics[scale=1,width=2\columnwidth]{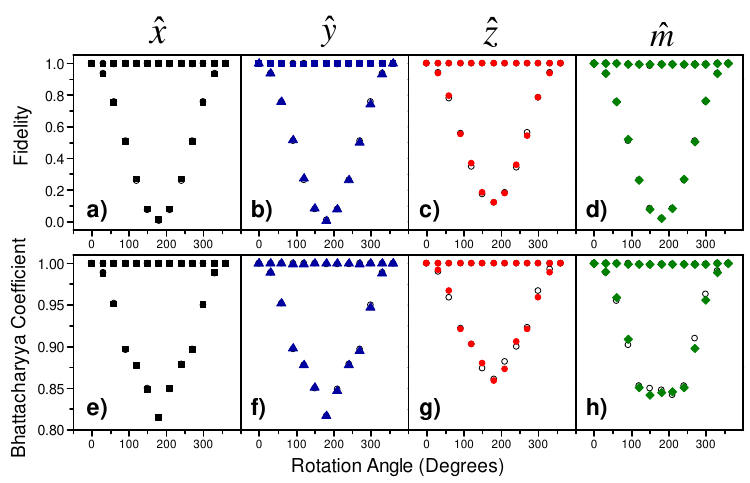}
  \caption{\label{fig:appendix}Summary of of the experimentally measured fidelity and
   the Bhattacharyya Coefficient for a wide range of unitaries. Panels
  a)--d) show the fidelity analysis results for unitary rotations
  about $\hat{x}$, $\hat{y}$, $\hat{z}$, and $\hat{m}$ axes as functions of rotation angle. The coloured data
  points are the comparison between stage \textbf{I} and stage \textbf{III}, and stage \textbf{I} and stage \textbf{II}. The open circles show
  the theoretical comparison which takes the state from stage \textbf{I} and applies theoretical
  unitaries. Panels e)--h) show the Bhattacharyya results comparing stage \textbf{I}
  and stage \textbf{III}, and stage \textbf{I} and stage \textbf{II}, in the coloured data points for each of the four axis, with
  the open circles being the theoretical comparison.}\label{fig:completeset}
\end{figure*}

Our experiment procedure included three stages, \textbf{I} measurements of the source,
 \textbf{II} measurements after we apply the unitary to only one qubit, and \textbf{III}
 measurements after applying the same unitary to both qubits.
The fidelities and Bhattacharyya Coefficients between stages
\textbf{I} and \textbf{II},
 and stages \textbf{I} and \textbf{III} as a function of the
 rotation angle are shown in Fig.~\ref{fig:appendix} for
 rotation axes, $\hat{x}$, $\hat{y}$, $\hat{z}$, and $\hat{m}$.
Panels a)--d) show the fidelity, and panels e)--h) show the
Bhattacharyya Coefficient ($BC$). The open circles show the
theoretical expectation for various unitaries. For the fidelity
comparison the theoretical model applies perfect unitaries to
the imperfect measured state. For the $BC$ comparison the
theoretical model applies perfect unitaries to the
reconstructed state from stage \textbf{I}. We observe very good
agreement between the measured and predicted results.

\newpage

\subsection{Fitting Son's theory to experimental data as a test of Born's rule}

In our experiment, we place a bound on the degree
of envariance. It has been shown that envariance can be used to derive Born's rule \cite{zurek03.1,born26}. However, the derivation
does not relate bounds on Born's rule to bound on envariance.
In order to do so, we explore a recently proposed extension of quantum mechanics
by Son \cite{wonmin}.  Son's theory generalizes Born's rule, replacing the 
familiar power of 2 which relates wavefunctions to probabilities with a power of $n$.
In this section, we summarize Son's theory and use it to put a bound on $n$
using our experimental data.

We first consider measurements on a pair of qubits in the maximally entangled singlet
state using standard quantum mechanics. We define measurement observables $\hat{a}=\vec{\alpha}\cdot\vec{\sigma_{1}}$
and $\hat{b}=\vec{\beta}\cdot\vec{\sigma_{2}}$ where $\vec{\alpha}$,
$\vec{\beta}$ are unit vectors and $\vec{\sigma_{1}}$, $\vec{\sigma_{2}}$
are the Pauli matrices for the two qubits. The result of measurements $a$ and $b$ for
qubits 1 and 2 respectively can take on the values $\pm1$. The correlation
function is defined by 
\begin{equation}
E=\langle ab\rangle=P_{a=b}-P_{a\neq b},
\end{equation}
where $P_{a=b}$ and $P_{a\neq b}$ are probabilities that
$a=b$ and $a\neq b$ respectively. The correlation
function only depends on the angle $2\theta$ between $\vec{\alpha}$ and $\vec{\beta}$ for the singlet state.
From Born's rule, we have the
probability amplitudes $\psi_{a=b}$ and $\psi_{a\neq b}$ satisfy
$P_{a=b}=|\psi_{a=b}|^{2}$ and $P_{a\neq b}=|\psi_{a\neq b}|^{2}$.
Therefore, the correlation function in standard quantum mechanics is given by
\begin{equation}
E_{QM}(\theta)=|\psi_{a=b}|^{2}-|\psi_{a\neq b}|^{2}=-\cos2\theta.
\end{equation}

\begin{figure}
  \centering
  \includegraphics[scale=1,width=1\columnwidth]{{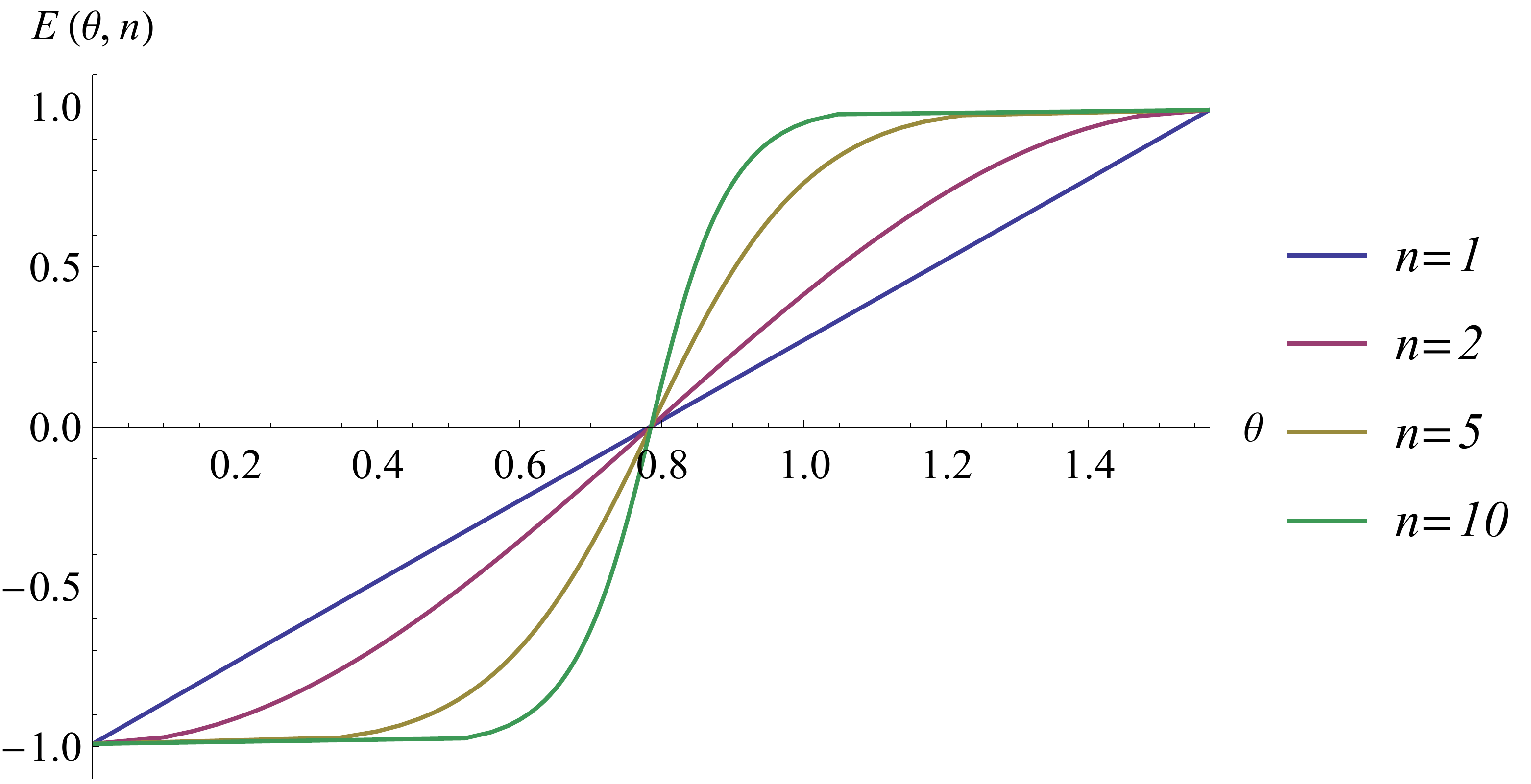}}
  \caption{Generalized correlations for the singlet state as a function of $n$ using Son's theory \cite{wonmin}. 
The correlation as a function of $\theta$ is shown for $n = 1$ (blue line), $n=2$ (purple line), $n =5$ (brown line) and $n=10$ (green line).  The $n=2$ case corresponds to standard quantum mechanics.}\label{fig:wonmin_theory}
\end{figure}

We now consider Son's theory, where Born's rule is generalized to be $P_{a=b}=|\psi_{a=b}|^{n}$
and $P_{a\neq b}=|\psi_{a\neq b}|^{n}$, and the correlation function is thus,
\begin{equation}
E(\theta,n)=|\psi_{a=b}|^{n}-|\psi_{a\neq b}|^{n},
\end{equation}
where standard quantum mechanics is the special case $E(\theta,2)=E_{QM}(\theta)$.
As in standard quantum mechanics, Son assumed that the correlation function depends only on the angle between measurement settings.
Son showed that the constraints $|\frac{\partial\psi_{a=b}}{\partial\theta}|^{2}+|\frac{\partial\psi_{a\neq b}}{\partial\theta}|^{2}\propto1$
and  $|\psi_{a=b}|^{n}+|\psi_{a\neq b}|^{n}=1$  and the boundary condition $E(0,n)=-1$ and  $E(\frac{\pi}{2},n)=1$ are sufficient to solve for $E(\theta,n)$. See \cite{wonmin}
for further details on the deviation. Figure ~\ref{fig:wonmin_theory} shows $E(\theta,n)$ for different value
$n$. 

In the experiment, we rotated one qubit while leaving the other qubit
unchanged during the stage \textbf{II} (See Figure ~\ref{fig:procedure}) . If we use the same measurement basis
on both qubits for that rotated state, we are effectively measuring
the singlet state input with two measurement basis with angle $\theta$
apart. For example, we can choose the rotation axis and the measurement basis to be [Z,(D,A)], where
the first qubit is rotated around Z axis, while measurements on the qubits are done in (D,A) basis.
Since the rotation axis Z is orthogonal to the measurement basis (D,A), we could view the rotation of qubit
as a rotation of the measurement basis in the D-A plane. For a rotation angle $\phi$, the angle between
two measurement basis is given by $2\theta=\pi-|\pi-2\phi|$. We could derive prediction of $E(\phi,n)$
from Son's theory, and test it with our data.

\begin{figure}
  \centering
  \includegraphics[scale=1,width=1\columnwidth]{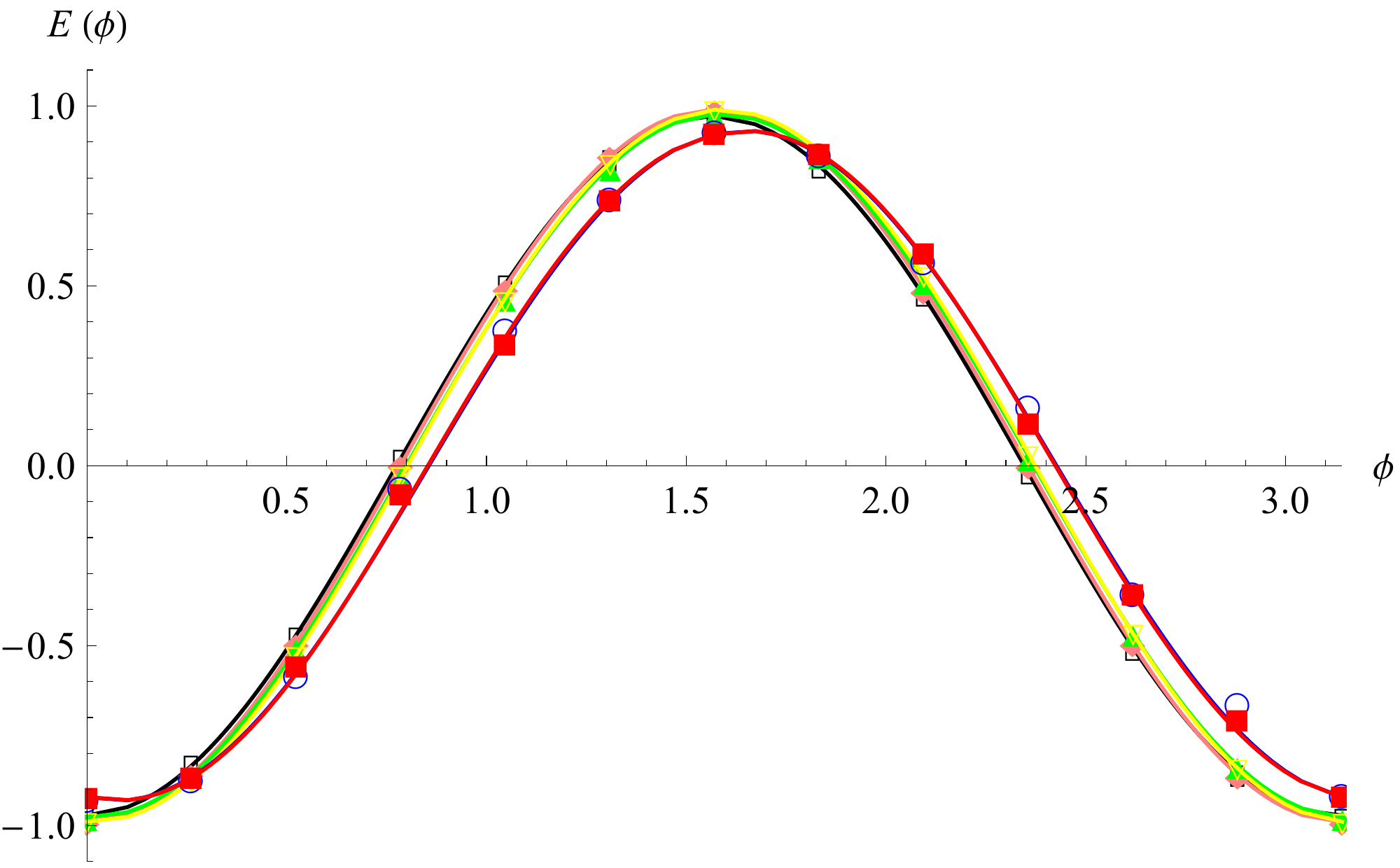}
  \caption{Correlation functions versus the rotation angle $\phi$. The experimental correlations are extracted from our data for the case where the rotation axis and the measurement basis are given by \{[Z,(D,A)], [Z,(R,L)], [Y,(D,A)], [Y,(H,V)], [X,(R,L)], [X,(H,V)]\} shown as \{red squares, blue circles, green up triangles, yellow down triangles, black empty squares, pink diamonds\} as a function of the rotation angle $\phi$. The best fit using Eq.~\ref{eq:SonThoeryModify} for each correlation is shown as a line whose colour matches the corresponding data points.  These fits yield estimates for the value of $n$ of \{2.04,
2.01, 2.00, 2.01, 2.01, 2.00\}, respectively }\label{fig:fit_plot}
\end{figure}

Son's derivation assumes a perfect singlet state which must be relaxed to obtain a comparison with experiment.  For a realistic state, the correlation function will not necessarily depend only on $\theta$.  In his derivation, Son additionally assumed $E(0,n)=-1$ and $E(\pi/2,n)=1$, i.e., perfect correlations, which are not experimentally achievable.  To relax these assumptions, we consider the difference between two correlation functions measured for a general state $\rho$ and the ideal state $|\psi^{-}\rangle$, $E(\phi,n, \rho)$ and $E(\phi,n,|\psi^{-}\rangle)$ where $\phi$ is the rotation angle of one of the settings.  For $n \approx 2$, we make the assumption that  $E(\phi,n,\rho)-E(\phi,n,|\psi^{-}\rangle)\approx E(\phi,2,\rho)-E(\phi,2,|\psi^{-}\rangle)$.  Thus for states close to the ideal singlet state and for $n$ close to 2, we have the relation:
\begin{equation}
E(\phi,n,\rho)\approx E(\phi,n,|\psi^{-}\rangle)+E(\phi,2,\rho)-E(\phi,2,|\psi^{-}\rangle).
\label{eq:SonThoeryModify}
\end{equation}
We calculated $E(\phi,2,\rho)$ and $E(\phi,2,|\psi^{-}\rangle)$
from standard quantum mechanics, and use Son's theory to calculate
$E(\phi,n,|\psi^{-}\rangle)$. For a given set of data $E_{exp}(\phi_{i})$,
we find $\rho$ and $n$ to minimize the objective function $L=\Sigma_{i}[E(\phi_{i},n,\rho)-E_{exp}(\phi_{i})]^2/[\delta E_{exp}(\phi_{i})]^2$,
where $\delta E_{exp}(\phi_{i})$ is the standard deviation of correlation function $E_{exp}(\phi_{i})$  predicted assuming Poissonian count statistics.
Figure~\ref{fig:fit_plot} shows the results of fitting the correlation functions for 6 sets of data.  From this, we extracted $n = 2.04, 2.01, 2.00, 2.01, 2.01, 2.00$; averaging these results and using their standard deviation to estimate the uncertainty yields $n=2.01 \pm 0.02$ in good agreement with Born's rule where $n = 2$.

\end{document}